\documentclass{article}
\usepackage{spconf,amsmath,graphicx}
\usepackage{url}
\usepackage{float}
\usepackage{amsmath}
\usepackage{array}
\usepackage{enumitem}

\title{Multitask learning and Multistage fusion \\ 
for Dimensional Audiovisual Emotion Recognition}
\name{Bagus Tris Atmaja$^{1,2}$, Masato Akagi$^1$}
\address{$^1$School of Information Science, Japan Advanced Institute of Science and 
Technology, Japan \\
$^2$Department of Engineering Physics, Sepuluh Nopember Institute of Technology, 
Surabaya, Indonesia \\
\url{{bagus, akagi}@jaist.ac.jp}}
\begin{document}

\maketitle
\begin{abstract}
Due to its ability to accurately predict emotional state using multimodal features, audiovisual emotion recognition has recently gained more interest from researchers. This paper proposes two methods to predict emotional attributes from audio and visual data using a multitask learning and a fusion strategy. First, multitask learning is employed by adjusting three parameters for each attribute to improve the recognition rate. Second, a multistage fusion is proposed to combine results from various modalities' final prediction. Our approach used multitask learning, employed at unimodal and early fusion methods, shows improvement over single-task learning with an average CCC score of 0.431 compared to 0.297. A multistage method, employed at the late fusion approach, significantly improved the agreement score between true and predicted values on the development set of data (from [0.537, 0.565, 0.083] to [0.68, 0.656, 0.443]) for arousal, valence, and liking.

\end{abstract}
\section{Introduction}
\label{sec:intro}

Automatic emotion recognition has been approached using two perspectives: 
the categorical view and the dimensional view. While most researchers attempted to 
categorize human emotion within different categories (e.g. happiness, anger, 
etc.), dimensional emotion recognition is the more challenging task as it seeks to 
label the emotions as degrees rather than as categories. 
From dimensional perspective, emotion is described in 2 or 3 
attributes \cite{russel1980three}. Valence (pleasantness) and arousal 
(emotion intensity) are the two most common dimensions in 2D emotion models. In 
3D models, either dominance (degree of control) or liking is used. Another model,
such as expectancy, can be added as a 4th dimension (4D) \cite{fontaine2007world}.

In this paper, we evaluated three emotional dimensions/attributes: arousal, valence, 
and liking, which have been obtained from the dataset in \cite{Ringeval19-A2W}. 
The task is to obtain the most accurate prediction on a specific metric. 
As a regression task, the most common 
metric is the error between true value and predicted emotion degree. However, 
recent researchers \cite{Ringeval19-A2W} introduced correlation measurement to 
determine the agreement between true value and predicted emotion degree.

Two approaches are commonly used to minimize the loss of learning 
process functionality and to obtain the best 
model to predict emotion dimension, i.e., 
single-task learning (STL) and multitask learning (MTL). Single-task learning 
minimizes single loss function only in multiple output learning. For example, 
when learning to predict arousal, valence, and liking in dimensional emotion, 
only arousal is minimized. The other dimensions, valence and 
liking, are ignored in the learning process. By minimizing the error of arousal, 
the result of the learning process can be used to predict one dimension 
(arousal) or all three dimensions (arousal, valence, and liking).

The problem with single-task learning is that, when it is used to predict multiple 
outputs, three scores are predicted using single loss function. 
A high score in one dimension usually resulted in a lower score on the other 
dimensions. To address this issue, 
we introduced the use of multitask learning when minimizing error between 
true emotion and predicted emotion degree for all emotion dimensions. 

The common approach in multitask learning is that the same weighting factors 
are used for each loss function in learning process. Therefore, the total is the sum of 
three each loss functions from each emotion dimensions. 
The method we propose in this paper is intended to obtain a balanced score by 
assigning a different weighting factor to each loss function for 
each emotion dimension. 

As emotion comes from many modalities, the creation of a fusion strategy to 
accommodate those modalities is also a challenge. 
The standard method is by combining the features among different modalities 
in either the same or different networks. This is called an early fusion 
strategy. Two or more feature sets are then trained to map those inputs onto 
labels. Another strategy is the use of late fusion. In this strategy, each 
modality is trained in its network using its label. The recognition 
results for each modality are then grouped to find the highest probability 
corresponding to the labels. The results from early fusion and late fusion also 
can be fused by combining those results in support vector regression (SVR). 
The result from this last step can be 
repeated in a multistage direction to improve the recognition rate.

Our contributions of this paper can be summarized as follows,
(1) the use of multitask learning to minimize
the loss function using three parameters for three emotion attributes from audiovisual features;
(2) the fusion strategy by analyzing unimodal and bimodal features on early 
fusion and late fusion, and combining early-late fusion using multistage 
SVR to improve audiovisual emotion recognition rate.

\section{Related Work}
\textbf{Multitask learning}. One of the problems in machine learning is to 
obtain the appropriate cost function or 
loss function to model the data. Most problems in regression analysis use error 
calculation between the true value and predicted value the loss function. 
The choice of the loss function is frequently determined by 
the metric used for evaluation. In the case of dimensional emotion 
recognition, Ringeval et al. of proposed the use of a concordance 
correlation coefficient (CCC) to score the performance of predicted emotion 
attributes \cite{Ringeval19-A2W}.

Parthasarathy and Busso used multitask learning to minimize mean 
squared error (MSE) in dimensional emotion recognition 
\cite{parthasarathy2017jointly}. The authors used two 
parameters to weigh loss function of three emotion attributes: arousal, 
valence, and dominance. Despite the weighting factor for both arousal and 
valence being determined, the weighting factor for dominance is obtained by 
subtracting 1 from the weighting factors of arousal and valence.  All weighting 
factors lie in a range of 0-1 with a 33.3\% possibility that one value is 
zero. It was also found that the best parameters are 0.7 and 0.3 for arousal and valence. 
In this case, dominance is ignored in learning process, which can be viewed as 
two-task learning which is similar to single task learning.

Using two approaches to multitask learning, such as shared layer and independent 
layer, the authors also achieved an improvement of CCC score 
compared to baseline single-task learning \cite{parthasarathy2017jointly}. 
As the system learned better on the larger network than on the smaller one,
the larger the network used, the greater improvement obtained,

Chen et al. also used multitask learning with MSE as the loss 
function \cite{chen2017multimodal}. Although the improvement of CCC score 
from the given baseline is achieved, 
the performance comparison to single-task learning is not specified. 
This potentially leads to a difficulty in determining whether the improvement came from 
multitask learning or other used strategies.

\noindent
\textbf{Multimodal Fusion.} As emotion can be recognized from many modalities, 
e.g., speech, facial image, movement, 
and linguistic information, the use of multimodal technique to accommodate 
many features is often considered in such systems. The dataset 
described in \cite{Ringeval19-A2W} includes multimodal emotion features from audio 
and visual. Busso et al. provided an emotion dataset from 
speech and gesture, including facial expressions and hand movements \cite{busso2008iemocap}. 
The improved version of that dataset provided an affective database with audiovisual information 
which promoting naturalness within the (acted) recording \cite{busso2016msp}. 

To deal with various features extracted from multimodal datasets, several 
categories of feature fusion have been developed by researchers 
\cite{chen2017multimodal, Zadeh2017, Majumder2018, Zhao2018, Atmaja2019b}. Most strategies can be 
divided into early fusion and late fusion. In an early fusion method, also 
known as feature level fusion, features from different modalities are combined 
before performing classification. In late fusion method, also known as decision 
level fusion, the final decision probabilities are given by each unimodal model 
results by such methods like SVR. 

Ringeval et al. provides baseline fusion method for late fusion 
strategy from SEWA dataset \cite{Ringeval19-A2W, kossaifi2019sewa}. The results from each 
modality or feature set can be combined using a static 
regressor, i.e., SVR to make the final decision of predicted emotion attribute 
scores from given results of several modalities.

\section{Data and Feature Sets}
The SEWA dataset \cite{kossaifi2019sewa} provided in \cite{Ringeval19-A2W} 
is used in this research. 
The dataset contains audiovisual recordings from: 
Chinese, English, German, Greek, Hungarian, and Serbian, but only 
German (DE) and Hungarian (HU) are used in this work as they did not 
provide test label in other languages. 
Three attributes provided to represent emotional states i.e.: arousal, valence, 
and liking. The scores of those attributes are obtained from annotation of 
several native speakers: six Germans and five Hungarians.
From 96 subjects, 68 subjects (34 each) are used in training, 
and the rest 28 subjects (14 each) are used for validation/development.

In addition to the dataset, the authors of paper \cite{Ringeval19-A2W} also provided 
baseline features which are shown in Table \ref{tab:feature_set}. Instead of 
generating new a feature set, we applied the
multitask learning and multimodal audiovisual fusion to those feature sets.

For both audio and visual features, the same processing blocks are used, i.e., 
4.0 s of window length and 100 ms of hop size, where the label is also 
given for each 0.1 s. The longest 1768 sequences (label numbers)  
is then used for all subjects by padding zeros for other sequences below this number. 
For bimodal feature fusion, audio and visual features are concatenated before 
they are fed into the classifier.

\begin{table}
\caption{Audio and visual feature sets evaluated in this research. Feature sets 
highlighted in bold are used for bimodal/multimodal emotion recognition.}
\begin{tabular}{l p{7cm}}
\hline
audio & eGeMAPS \cite{Eyben,Eyben2013}, 
        \textbf{Bag-of-Audio-Word eGeMAPS (BoAW-e)} \cite{Schmitt2016}, 
        \textbf{eGeMAPS functional}, \textbf{DeepSpectrum} 
        (DS) \cite{amiriparian2017snore}, MFCCs, \textbf{BoAW MFCCs (BoAW-M}) 
        \cite{Schmitt2016}, 
        MFCCs functionals.\\
\hline
visual & Facial Activation Units (FAUs) \cite{baltrusaitis2018openface}, 

         \textbf{FAUs functionals}, 
         \textbf{ResNet} \cite{simonyan2014very}, VGG  \cite{he2016deep}.\\
\hline
\end{tabular}
\label{tab:feature_set}
\end{table}

\section{Proposed Method}
\subsection{Multitask learning based on CCC loss}
\begin{figure}
    \centering
    \includegraphics[width=3.2in]{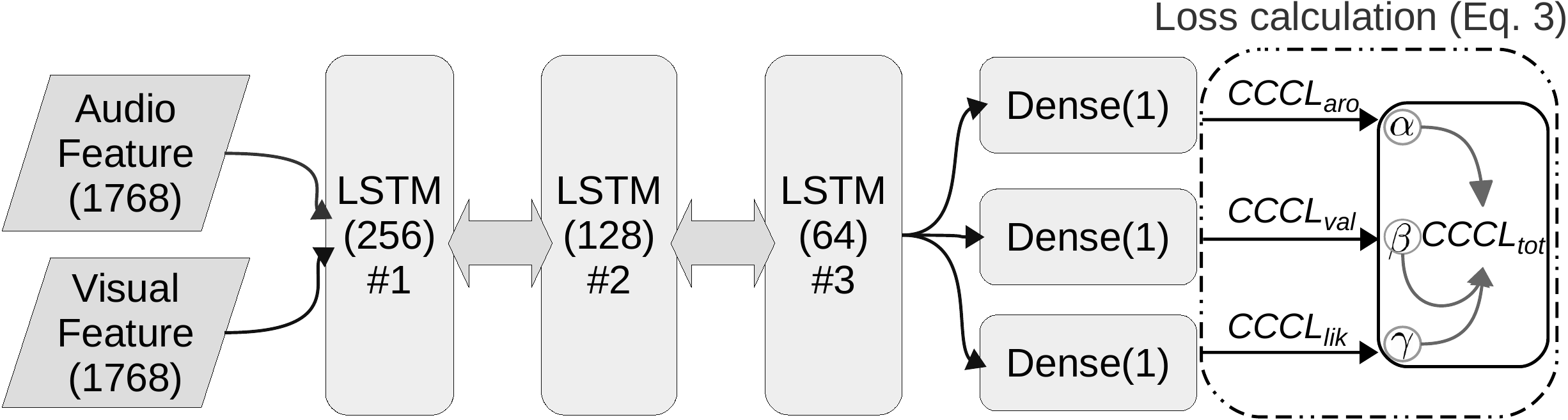}
    \caption{The architecture of Deep Neural Network (DNN) with a multitask 
    learning approach for minimizing loss function from three dense layers. 
    The number inside the bracket represents units.
    }
    \label{fig:system}
\end{figure}




CCC is the common metric in dimensional 
emotion recognition to measure the agreement between true emotion dimension with 
predicted emotion degree. The CCC is formulated
\begin{equation}
    CCC = \dfrac {2\rho_{xy} \sigma_{x} \sigma_{y}}
        {{\sigma_{x}^2}+\sigma_{y}^2 + (\mu_x - \mu_y)^2},
\end{equation}
where $\rho_{xy}$ is the Pearson coefficient correlation between $x$ and $y$, 
$\sigma$ is standard deviation, and $\mu$ is a mean value. 
This CCC is based on Lin's calculation \cite{lawrence1989concordance}. 
The range of CCC is from $-1$ (perfect disagreement) to $1$ (perfect agreement). 
Therefore, the CCC loss function (CCCL) to maximize the agreement between true 
value and prediction emotion can be defined as 
\begin{equation}
    CCCL = 1 - CCC
\end{equation}
In single-task learning, the loss function is one of the loss functions 
from arousal ($CCCL_{aro}$), valence ($CCCL_{val}$), or liking ($CCCL_{lik}$).
In multitask learning, when CCC loss is used as a single metric for  all 
arousal, valence, and liking, the $CCCL_{total}$ is a combination of those three CCC 
loss functions: 
\begin{equation}
    CCCL_{tot} = \alpha ~ CCCL_{aro} + \beta ~ CCCL_{val} + \gamma ~ CCCL_{lik},
\end{equation}
where $\alpha$, $\beta$, and $\gamma$ are the weighting factors for each emotion 
dimension loss function. In a common approach, $\alpha$, $\beta$, and $\gamma$ 
are set to be 1, while in \cite{parthasarathy2017jointly}, $\gamma$ is set to 
be $1-(\alpha + \beta)$ to minimize MSE. In that approach, all weighting 
factors are in range 0-1.

In this paper, we use all three parameters, and the sum of those 
weighting factors is not limited to only 0-1. As the goal is to strengthen 
CCC, CCC loss is used instead of MSE.

As shown in Fig. \ref{fig:system}, the audiovisual emotion recognition system 
consists of 3 LSTM layers with 256, 128, and 64 units. A dropout layer with 
a factor of 0.4 is added after each LSTM layer. A RMSprop optimizer is used 
with a learning rate of 0.0005 and 34 batch size for 50 epochs in one experiment. 
To compensate for the delay when making an annotation, the label is shifted 0.1 to the 
front in the training process and shifted back in writing the prediction.

\subsection{Multistage Fusion using SVR}
\begin{figure}
    \centering
    \includegraphics[width=3.3in]{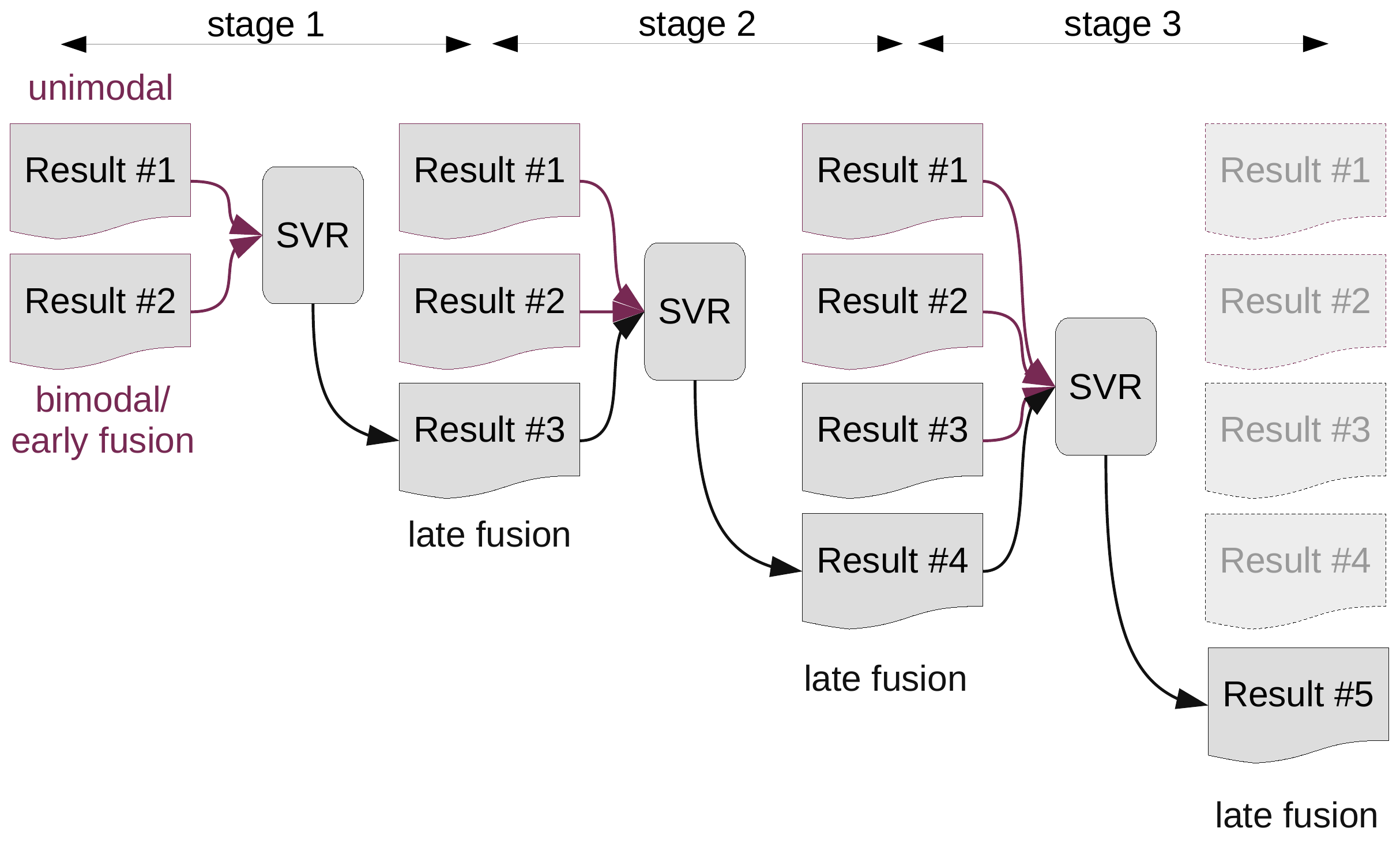}
    \caption{The flow of multistage SVR to combine early and late fusion results.}
    \label{fig:early_late}
\end{figure}

In Fig. \ref{fig:system}, the system produces a prediction of arousal, valence, 
and liking degree from bimodal audio and visual feature sets. This result 
can be combined with the other results from the unimodal or bimodal (early) fusion 
using SVR (from different feature set), and the resulting prediction from SVR 
also can be input to the same SVR system (implemented using scikit-learn tool 
\cite{scikit-learn}). In Figure 
\ref{fig:early_late}, this combination of early fusion and late fusion is 
illustrated in three stages. First, the result from unimodal, named as 
result \#1, and multimodal (bimodal), named as result \#2, or unimodal and 
unimodal are trained using SVR method. This 
learning process results in a new result (namely, result \#3 in that Figure). 
The result \#3 from late fusion is fed again to SVR method results in result \#4. 
Result \#4 is fed again to SVR method results in result \#5. This multistage 
fusion can be performed $n$-times to gain improvement of CCC score.

\section{Multitask Learning Results}
To evaluate the effectiveness of the proposed MTL method versus STL and previous MTL 
methods, we compared CCC scores among those methods. Table \ref{tab:mtl_compare} 
shows CCC scores for different attributes with its average. Our 
proposed MTL2 outperforms STL and previous proposed MTL1. To find the optimum 
parameter of $\alpha$, $\beta$, and $\gamma$, we performed random search for 
those parameters in range 0-1. The parameters used in Table \ref{tab:mtl_compare} 
are the optimum ones i.e. 0.7, 0.2, and 1.0 for $\alpha$, $\beta$, and $\gamma$, 
respectively.

Our proposed MTL learning with three parameters outperforms STL 
and previous MTL \cite{parthasarathy2017jointly}. For STL approaches, both arousal and 
valence obtained the highest CCC score when its attribute is optimized. 
Although the liking is optimized in STL3, it remains the most difficult to estimate.
This problem should be addressed in future research.

\begin{table}
\caption{CCC score of development set from FAUs feature set comparing STL and MTL. 
STL is performed by setting a weighting factor to 1 for the related attribute 
(Eq. (3)).}
\begin{tabular}{l c c c c}
\hline
Loss        & Arousal   & Valence   & Liking   & Average\\
\hline
STL1 (Aro)  & 0.511 & 0.235 & 0.107 & 0.284\\
STL2 (Val)  & 0.255 & 0.558 & 0.077 & 0.297\\
STL3 (Lik)  & 0.255 & 0.32  & 0.191 & 0.244\\
MTL1 \cite{parthasarathy2017jointly} & 0.476 & 0.524 & 0.009 & 0.336\\
MTL2 (ours) & \textbf{0.522} & \textbf{0.578} & \textbf{0.194} & \textbf{0.431}\\
\hline
\end{tabular}
\label{tab:mtl_compare}
\end{table}

%

\section{Multistage Fusion Results}
To obtain multistage fusion results, the following steps are performed,
\begin{enumerate}[nosep]
\item Unimodal emotion recognition: This step is performed to investigate 
      the importance feature set for bimodal or multimodal fusion.
\item Bimodal fusion: This step is performed by concatenating two feature sets, 
      from different or same modality.
\item Multimodal fusion: While the first two steps are performed using DNN, this 
      third step is performed using SVR by combining results from unimodal or 
      bimodal emotion recognition.
\item Multistage fusion: Output from multimodal SVR can be combined using 
      the same SVR to improve the recognition rate of emotion recognition.
\end{enumerate}
We run experiments on unimodal feature sets by inputting one feature set into 
a system to find which feature sets gives better performance. 
For this purpose, we use small networks with previously explained LSTM 
layers (implemented in Keras \cite{chollet2015keras}). From 12 feature sets, we 
choose 7 feature sets by highest average CCC scores. The combination of 
seven feature sets resulted in 21 pairs of bimodal feature sets. Note 
that the definition of bimodal here is not audio and visual modalities but 
a pair of two feature sets. From unimodal and bimodal results, we choose the 11 
highest CCC scores and input those 11 results to SVR to perform multimodal 
audiovisual emotion recognition by late fusion.

\begin{figure}[!htb]
\centering
\includegraphics[width=3.4in]{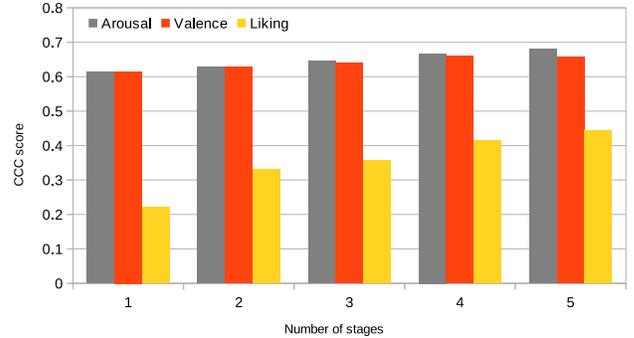}
\caption{CCC score among attributes from the different numbers of stages in 
proposed multistage feature fusion using SVR.}
\label{fig:multistage_result}
\end{figure}

This last multimodal fusion using SVR can be regarded as 1-stage feature 
fusion. By inputting the result from SVR to the same SVR system,
a 2-stage multimodal fusion can be performed. We limited this multistage 
multimodal fusion to 5 repetitions. The result of CCC scores for arousal, valence, 
and liking from 1 to 5 stages is shown in Fig. \ref{fig:multistage_result}. 
That figure shows that CCC scores improved as the number of stages increased.

In comparison with unimodal, bimodal, and multimodal fusion (1 stage), multistage 
fusion gained significant improvements. The proposed multistage fusion 
could improve CCC score of liking attribute, which is the most challenging 
attribute in this task, from 0.083 (baseline unimodal) to 0.443. Other two 
attributes obtained relative improvement over baseline results of 26.63\% and 16.11\% respectively for 
arousal and valence.

\begin{table}
\caption{CCC score comparison of development set on bilingual dataset by different
methods; each row is the highest obtained score among feature sets using the same method.}
\begin{tabular}{l c c c}
\hline
Method  & \multicolumn{3}{c}{CCC Development (DE+HU)} \\    
        & Arousal    & Valence   & Liking    \\
\hline
Baseline (FAUs) \cite{Ringeval19-A2W}     &   0.531    &  0.565    &   0.083  \\
Unimodal            &   0.522    &  0.578  &   0.194    \\
Bimodal early fusion  &  0.552 &  0.557 &    0.284    \\
Multimodal late fusion&   0.627    &  0.616    &   0.292  \\
Multistage Fusion   &   \textbf{0.680}    &  0.\textbf{656}    &   \textbf{0.443}\\
\hline
\end{tabular} 
\label{tab:test_score}
\end{table}

\section{Conclusions}
A multitask learning strategy is proposed to balance the CCC score among arousal, valence, and liking 
by adjusting parameters for those attributes. 
The result shows that by using different weighting factors for each emotional dimension, 
an improvement in terms of CCC scores can be obtained. Using weighting factors of
0.7, 0.2, and 1.0 for arousal, valence, and liking, respectively, for MTL parameters,
we achieved an improvement of average CCC score from 0.297 using STL to 0.431 using 
our MTL. To deal with multimodal fusion of several feature sets, we proposed 
a multistage fusion using SVR method. This proposed method improves the CCC score 
on development test significantly for bilingual emotion recognition (DE+HU), 
especially on liking attribute i.e., [0.680, 0.656, 0.443] with an average CCC 
score of 0.593.

\bibliographystyle{IEEEbib}
\bibliography{bta_icassp2020}

\end{document}